\documentclass[twocolumn,showpacs,preprintnumbers,amsmath,amssymb]{revtex4}

\usepackage{graphicx}
\usepackage{dcolumn}
\usepackage{bm}
\usepackage{mathtools}

\DeclareMathAlphabet{\mathpzc}{OT1}{pzc}{m}{it}

\begin{document}
	
\title{Noncommutative black hole in the Finslerian spacetime}
	
\author{Sourav Roy Chowdhury}
\email{sourav.rs2016@physics.iiests.ac.in}
\affiliation{Department of Physics, Indian Institute of Engineering Science and Technology,
		Shibpur, Howrah 711103, West Bengal, India}
	
\author{Debabrata Deb}
\email{ddeb.rs2016@physics.iiests.ac.in}
\affiliation{Department of Physics, Indian Institute of Engineering Science and Technology,
		Shibpur, Howrah 711103, West Bengal, India}
	
\author{Farook Rahaman}
\email{rahaman@associates.iucaa.in}
\affiliation{Department of Mathematics, Jadavpur University, Kolkata 700032, West Bengal, India}
	
\author{Saibal Ray}
\email{saibal@associates.iucaa.in}
\affiliation{Department of Physics, Government College of Engineering and Ceramic Technology, Kolkata 700010, West Bengal, India  
		\& Department of Natural Sciences, Maulana Abul Kalam Azad University of Technology, Haringhata 741249, West Bengal, India}
	
\author{B.K. Guha}
\email{bkg@physics.iiests.ac.in}
\affiliation{Department of Physics, Indian Institute of Engineering Science and Technology,
		Shibpur, Howrah 711103, West Bengal, India}
	
\date{\today}
	
\begin{abstract}
We study the behavior of the noncommutative radiating Schwarzschild black hole in the Finslerian spacetime. The investigation shows that black hole possesses either (i) two horizons, or (ii) a single horizon, or (iii) no horizon corresponding to a minimal mass. We obtain that the minimal mass significantly changes with the Finslerian parameter, keeping minimal horizon remain unchanged. It turns out that under Finslerian spacetime, the maximum temperature before cooling down to absolute zero varies with Finslerian parameter. We then study the stability of the black hole by analyzing the specific heat and free energy. The energy conditions, their violation limit also scrutinized. Our findings suggest a stable black hole remnant, whose mass and size are uniquely determined in terms of the  Finslerian parameter $\overline{Ric}$ and noncommutative parameter $\theta$. The physical relevance of these results are discussed in a brief.
\end{abstract}
	
\pacs{02.40.Gh, 04.40.Dg, 04.20.Jb, 04.70.Dy}
	
\maketitle

\section{INTRODUCTION} 
In classical general relativity, the notion of the black hole is as an object in spacetime, exhibiting such strong gravitational acceleration that no particles even electromagnetic radiation can escape from it. The underlying nature and behavior of such objects have been targets of many theoretical researchers. In the curved background, based on quantum field theory, Hawking has been shown that a black hole can evaporate by emitting thermal radiation analogous with the black body radiation~\cite{hawking1,hawking2}. The discovery of Hawking radiation solved the problem in black hole thermodynamics and reconciliation of the quantum mechanics and gravity. Hawking has pointed out that the virtual particles with negative energy in the vicinity of the black hole can come into the black hole through the tunneling effects. As a consequence, the energy of the system and the radius of the event horizon will decrease. One of the important properties of a black hole is Hawking temperature, which is directly related to the surface gravity of the black hole on the horizon. Bekenstein proposed that the entropy of the black hole is also proportional to the area of the horizon~\cite{Bekenstein1,Bekenstein2}. The entropy of a black hole is maximum in comparison to any object of the same size. 
	
The final state of black holes evaporation issue can settle down in possible ways. One of these is as Hawking suggested, no remnant leaving behind the complete evaporation of black hole except only incoherent radiation. As a  consequence, loss of quantum coherence does occur. Alternatively, stable remnants leave behind the Hawking process and in the initial configuration of the system which reflects in precise quantum states~\cite{Aharonov}. To define this, the number of distinct virtual quantum states are needed. There are two different aspects to be accounted for physical consideration. The first issue is the remnant's size and mass in the order of Planck scale, i.e. infinite density of states. If gravity in any regular way coupled with these light remnants, in the presence of the energy of the Planck Scale amount, there would be profuse production in pairs. Another point is that the Hawking process might terminate, in spite of the black holes still having macroscopic mass. That could only happen if a singularity develops on the apparent horizon of the geometry by the back reaction~\cite{Susskind1}. According to the string/black hole correspondence principle~\cite{Susskind2}, at the late stage of the black hole the stringy effects cannot be neglected. Out of different conclusions of string theory, we are interested in the result in which the spacetime coordinates itself become the noncommutating operators on a D-brane~\cite{Witten,Seiberg}.
	
Usually considered spacetime continuum does not require any Lorentz invariance. The idea of Lorentz invariant discrete spacetime, i.e. spacetime quantization, long ago provided by Snyder~\cite{Snyder}. The noncommutative behavior of the spacetime is encoded by the relation
	\begin{equation}
	[\textbf{x}^\mu, \textbf{x}^\nu ]=i\theta^{\mu \nu},\label{1}
	\end{equation}
where $\theta^{\mu \nu}$ is an anti-symmetric matrix. It determine the discretization of the cell of spacetime in the same manner as the phase space discretized by the Planck constant $\hslash$. The noncommutativity is an intrinsic parameter of spacetime and independent of the behavior of curvature~\cite{Gruppuso,Nicolini1}. It eliminates the point like pattern and can be replaced with smeared objects. The modified quantum field theory (QFT), after implementation of the commutator relation Eq.(\ref{1}) has been largely investigated based on two distinct approaches: (i) Wely-Winger-Moyal *-product and (ii) coordinate coherent state formalism~\cite{Smailagic1,Smailagic2}. It has been shown that the doubts raised on the Lorentz invariance and unitary~\cite{Chaichian,Cho} in the *- product approach can figure out on considering $\theta^{\mu \nu}~=~\theta$~diag($\epsilon_1,...\epsilon_{D/2}$)~\cite{Smailagic3} where $\theta$ is a constant of the dimension of the order length square.
	
To incorporate the effects of noncommutativity, one could think of to modify the $r-t$ section of the metric of spherical symmetry and as a result the modified form of the Einstein field equations. However, Nicoloni et al.~\cite{Nicolini2} argued that this is not necessary. Retaining the tensor part same of the field equation, one can only modify the matter source of the system. The geometrical structure over the manifold is defined by the metric field and the strength of the field is measured by the curvature, i.e. the response of the existence of the matter distribution. Details study of the influence of noncommutativity in general relativity (GR) and the matter, energy and momentum distribution and propagation has been studied by several scientists~\cite{Smailagic1,Smailagic2,Smailagic3}. For the regular black hole, the quantum cooling process of evaporation has been studied in Ref.~\cite{Myung} whereas thermodynamical behavior of the noncommutative black hole has been investigated by Banerjee et al.~\cite{Banerjee1}. They provided the generalized form of black hole temperature and surface gravity, and shown that the relation is valid until quantum relation is negligible. In a subsequent study~\cite{Banerjee2}, Banerjee et al. provide the corrected form of area law in noncommutative geometry. 

In semiclassical form, Garattini and Lobo derived the exact solutions for worm holes~\cite{Garattini}. Later, the exact solution and physical characteristics have been explored~\cite{Lobo}. It is found that in the noncommutative framework the special class of thin wormholes, which are unstable in GR, are stable enough to small linearized radial perturbation~\cite{Kuhfittig}. Study of higher dimensional wormholes is available in Ref.~\cite{Rahaman3}. In (2+1)-dimensions Ba\~{n}dos-Teitelboim-Zanelli (BTZ) black hole constructed and corresponding exact solution in the framework of noncommutative geometry has been studied in Ref.~\cite{Rahaman1}. Due to uncertainty encoded in the commutative coordinate, the energy density can diffuse. In the context of noncommutative geometry the gravitational rotational curve can easily be explained. Noncommutativity is sufficient to produce stable circular form, without the help of dark matter~\cite{Rahaman2}. In order to get the expression of energy, the energy-momentum for a noncommutative radiating black hole has been considered with both the Einstein and M$\varnothing$ller prescription~\cite{Radinschi}. The result shows that the Einstein platform is more powerful. Kim et al.~\cite{Kim} provide a comparative study of the similarity of thermodynamic relations between the noncommutative Schwarzschild black hole and the Reissner-Nordstr\"{o}m black hole.
	
The notion of Finsler geometry based on the time measurement between two events that passes an observer is equivalent to the length which connects the events along the observer's world line. The quantification is on the tangent bundle of a homogeneous function~\cite{Bao2000,Pfeifer2014}. The geometrodynamics can be explained, apart from curvature, with the color property of the manifold.  It brings positional and directional dependent behavior, along with the intrinsic local anisotropy~\cite{Kouretsis2012}. 
	
Riemann introduced a metric structure, based on arc element in general space as
	\begin{equation*}
	x ds = \mathcal{F}(x_1,~ : ~:~ x_n; y_1,~ : ~:~ y_n),
	\end{equation*}
where $\mathcal{F}(x; y)$ is a positive homogeneous function of degree one in $y$. Here, $x$ and $y$ representing the position vector and tangent vector respectively. However, without any the quadratic restriction, the easiest description of Riemannian geometry is in Ref.~\cite{Bao2000} as follows 
	\begin{equation*}
	\mathcal{F}^2=g_{\mu \nu} y^i y^j.
	\end{equation*}
	
In Finslerian geometry, the metric geometry (including the Lorentz metric) can be extrapolated by defining a length for the curve. Instead of the metric, a general length measurement for curves on $ \mathbb{M}$ derives the geometry. The whole concept was primarily proposed by Finsler~\cite{Finsler1918}. The geometry is engaged to matter dynamics and on assumption that the metric tensor is independent of dynamics, the system reduces to Riemannian.
	
A complete impact of a Finslerian modification of spacetime in the astrophysical range is still missing. As a possible alternative spacetime model, Finslerian modification is already considering in astrophysics, cosmology, GR and different gravities. Several similarities between the Finsler geometry of Ransder space and Conformal geometry in classical spacetime are available in the Refs.~\cite{Caponio1,Caponio2,Caponio3}. Timelike and null geodesics as well as the geodesic motion of particles on the cosmologically symmetric spacetime studied by Hohmann~\cite{Hohmann2017}. The notion of Finsler geometry is developed from quantum gravity ideas~\cite{Girelli2007}. Later on, in special relativity, it is testified by Gibbons~\cite{Gibbons2007}. Singularity theorem and the Raychaudhuri equation can be obtained in Refs.~\cite{Minguzzi2015,Stavrinos2018}. The singularity theorem Penrose has been obtained in Ref.~\cite{Aazami}. In this paper, we specifically would like to investigate the characteristics and behavior of the black hole inspired by noncommutative geometry in the framework of Finsler spacetime. The motivation of considering Finsler geometry as an alternative model of spacetime is, no quadratic restriction and arc length is not only the function of length but velocity also. The field equations are obtained by considering the flag curvature. For simplicity, we consider the constant flag behavior, which admits spherical symmetric behavior.

\section{The interior structure}
We would like to analyze the Schwarzschild black hole inspired by the noncommutative geometry in the framework of Finslerian spacetime. Therefore, let us consider the line element to describe the static, spherically symmetric interior spacetime is in the form
	\begin{equation}
	\mathcal{F}^2= -e^{\nu(r)} y^t y^t + e^{\lambda(r)} y^r y^r +r^2 \overline{\mathcal{F}}^2 (\theta, \phi, y^\theta, y^\phi ), \label{2}
	\end{equation}  
where  $\mathcal{F} = \mathcal{F}(x, y)$ is the Finsler metric on a manifold $\mathbb{M}$, is a function of $(x^\mu, y^\mu)$ in a standard coordinate system.
	
Let $\mathbb{K}$ be a standard killing vector field  and  a smooth diffeomorphism $f:\mathbb{R} \times \mathbb{M}^\prime \rightarrow \tilde{\mathbb{M}}$, such that $f*\mathbb{K} = \partial_t$ and the line element $L(f_*(\tau,v))$. Let assume $L':= Lof_*$. $L$ is continuous on $\mathbb{TM}$. Though, not smooth Since it lacks twice differentiability, as $L'$ is not along $\partial_t$. This is possible iff in $\tilde{\mathbb{M}}$, both $\mathbb{K}$ and $\mathbb{T}$ are collinear at every point. The system follows semi Riemannian system as both Killing vector fields for $\tilde{g}$ are in same timelike cone. Hence, they are proportional, as follows: $\mathcal{L}_\mathbb{T} \tilde{g}=\mathcal{L}_\mathbb{K} \tilde{g}=0$. Details study for the standard static spacetime has been considered in Ref.~\cite{Caponio4,Caponio5}. 
	
For static and spherically symmetric metric, the energy-momentum tensor is given by $T^{\mu}_{\nu}$ = diag$(-\rho_{\theta}, p_r, p_t, p_t)$. Corresponding Einstein field equations can be write as
	\begin{align} \centering
	\frac{\lambda' e^{-\lambda}}{r} - \frac{ e^{-\lambda}}{r^2} +\frac{\overline{Ric}}{r^2} &= 8 \pi_\mathcal{F} \rho_{\theta}, \label{3}\\
	\frac{\nu' e^{-\lambda}}{r} + \frac{ e^{-\lambda}}{r^2} - \frac{\overline{Ric}}{r^2} &= 8 \pi_\mathcal{F} p_r, \label{4}\\
	e^{-\lambda}\left[ \frac{\nu''}{2}+\frac{\nu'^2}{4}-\frac{\nu' \lambda'}{4} +\frac{\nu'-\lambda'}{2r} \right] &= 8 \pi_\mathcal{F} p_t, \label{5} 
	\end{align}
where $\overline{Ric}$ represents the Ricci scalar, derived from $\overline {\mathcal{F}}^2$.
	
In flat spacetime, point-like structures are eliminated by noncommutativity and replace with smeared objects. Mathematically smearing can be implemented as a position Dirac-delta function replaced by Gaussian distribution of minimal width $\sqrt{\theta}$. In connection to that, we choose the mass density for the anisotropic fluid distribution as
	\begin{equation}
	\rho_{\theta}=\frac{M}{(4 \pi \theta)^{3/2}}\exp (-r^2/4\theta)=-p_r,\label{6}
	\end{equation}
where the total mass ($M$) of the source is diffused over $\sqrt{\theta}$ sized linear region. Minimal deviation from standard vacuum Schwarzschild geometry can be expected at a large distance, as well as in the range r $\simeq \sqrt{\theta}$.
	
\begin{figure}[htp!]
		\includegraphics[scale=0.35]{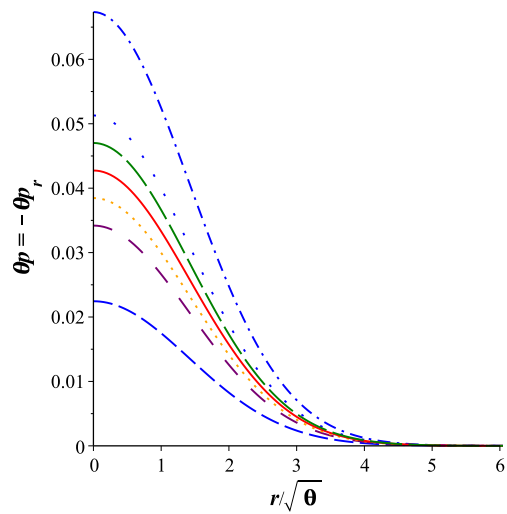}
		\caption{Variation of $\theta_{\rho}=-\theta_{p_r}$vs $r\sqrt{\theta}$. Line dot for M=$\sqrt{\theta}$, Space line for M=3$\sqrt{\theta}$, Long dot stands: M=2.285$\sqrt{\theta}$, Long dash stands: M=2.094$\sqrt{\theta}$, Solid line stands: M=1.905$\sqrt{\theta}$, Dot stands: M=1.713$\sqrt{\theta}$, Space dash stands: M=1.522$\sqrt{\theta}$; Blue stands: $\overline{Ric}=1.2$, Green stands: $\overline{Ric}=1.1$, Red stands: $\overline{Ric}=1.0$, Orange stands: $\overline{Ric}=0.9$, Burgundy stands: $\overline{Ric}=0.8$. Hereafter, color and line style maintain the corresponding $\overline{Ric}$ and mass.}\label{frho}
	\end{figure}
	
Consider the variation of matter density in Fig.~\ref{frho}. In the vicinity of origin ($r<<\sqrt{\theta}$), the variation of matter density is almost flat, i.e. $\frac{d\rho_{\theta}}{dr}\simeq ~0$. Again far away from the origin ($r>>4\sqrt{\theta}$), the variation is also flat; $\rho_{\theta}(0)>>\rho_{\theta}(r')$. The central density ($\rho_{\theta}(0)$) is higher for the higher total mass.
	
We have the conservation equation as follows:
	\begin{equation}
	(\rho_{\theta}+p_r)\frac{\nu'}{2}+p_r'-\frac{2}{r}(p_t-p_r)=0.\label{7}
	\end{equation}
	
For physically validity, the inward pull must balanced by the non-vanishing radial pressure. This is the effect by the noncommutative spacetime on the matter at a distance scale of order $\sqrt{\theta}$.

From relation Eq. (\ref{6}) and (\ref{7}), we obtain 
	\begin{equation}
	p_t = \frac{M}{(4 \pi \theta)^{3/2}}\Big(\frac{r^2}{4 \theta}-1\Big)e^{-r^2/4\theta}.\label{8}
	\end{equation}

\begin{figure}[htp!]
		\includegraphics[scale=0.35]{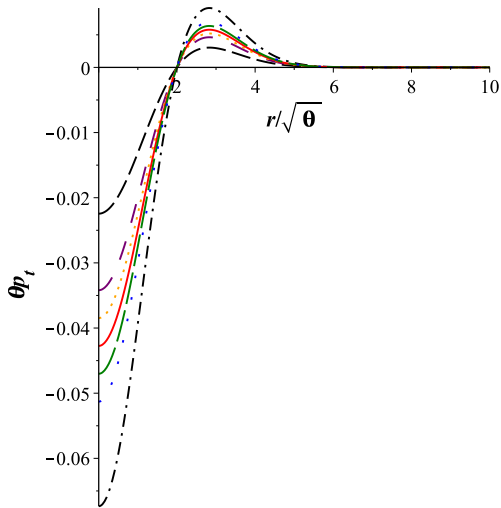}
		\caption{Variation of $\theta_{p_t}$ vs $r\sqrt{\theta}$ for different values of $\overline{Ric}$. }
		\label{fpt}
	\end{figure}

Variation shown in Fig.~\ref{fpt}.
	
We obtain the following metric coefficients for the matter distribution provided in Eq. (\ref{6}) for the line element in Eq. (\ref{2})
	\begin{eqnarray}
	& &e^{\nu} = 1 -\frac{4M}{r \sqrt{\pi}\overline{Ric}} \gamma \Big(\frac{3}{2}, \frac{r^2}{4 \theta}\Big),\label{9}\\
	& &e^{-\lambda} = \overline{Ric} -\frac{4M}{r \sqrt{\pi}} \gamma \Big(\frac{3}{2}, \frac{r^2}{4 \theta}\Big),\label{10}
	\end{eqnarray}
where $ \overline{Ric} $ represents the Ricci scalar, derived from $ \overline{F}^2 $ and  $\gamma \Big(\frac{3}{2}, \frac{r^2}{4 \theta}\Big)$ is the lower incomplete gamma function (provided in Eq. \ref{eq1}).

	\begin{figure}[htp!]
		\includegraphics[scale=0.35]{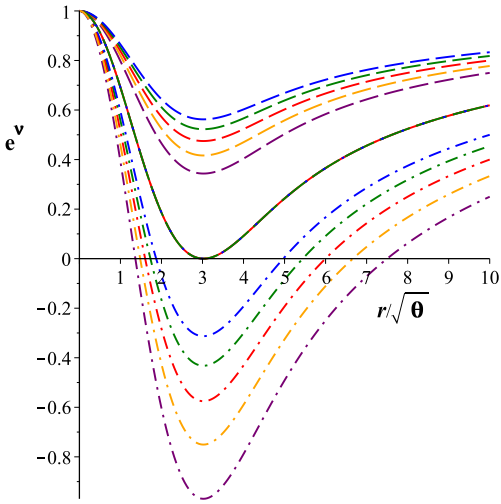}
		\caption{Variation of $e^{\nu}$ vs $r\sqrt{\theta}$ for different values of $\overline{Ric}$.}
		\label{fnu}
	\end{figure}

	\begin{figure}[htp!]
		\includegraphics[scale=0.35]{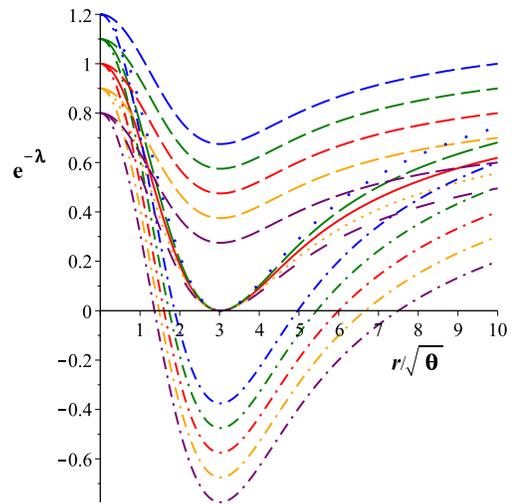}
		\caption{Variation of $e^{-\lambda}$ vs $r\sqrt{\theta}$ for different values of $\overline{Ric}$. }
		\label{flam}
	\end{figure}
	
For $\theta\rightarrow0$, the commutative Schwarzschild metric is obtained from the noncommutative metric. The exterior Schwarzschild metric is provided in Ref.~\cite{Li2014}. The variation of metric potentials $e^{\nu}$ and $e^{-\lambda}$ are shown in Figs. \ref{fnu} and \ref{flam} respectively. From plots it is clear that there are two horizons, (i) the inner horizon (Cauchy horizon) $r_c$ and (ii) the outer horizon (event horizon) $r_H$.  There exists a minimal mass $M_o$ below which no horizon, i.e. no black hole can form (space dot line). Again two distinct horizon formed for $M>>M_o$ (dash line). For $M=M_o$, the inner and outer both horizons met, and one degenerate horizon (minimal horizon) found at that point ($r_c\leq r_o\leq r_H$). It is found from the plots that for the varying $\overline{Ric}$, the degenerate horizon remain the same, though the minimal mass changes with $\overline{Ric}$. A higher $\overline{Ric}$ is needed for the higher minimal mass to form a single degenerate event horizon. 
	
The mass distribution of the system can directly be obtained with the help of Eq. (\ref{6}), as follows:
	\begin{equation}
	m(r) = \frac{2M}{ \sqrt{\pi}} \gamma\Big(\frac{3}{2}, \frac{r^2}{4 \theta}\Big).\label{11}
	\end{equation}

\section{Characteristics of black hole}
In this Section, we study few characteristics and stability of the black hole.
	
The event horizon of the black hole can be found either from Eq. (\ref{9}) or (\ref{10}), by considering $e^{\nu}$=0=$e^{-\lambda}$ in the following form
	\begin{equation}
	r_H = \frac{4M}{ \sqrt{\pi}\overline{Ric}} \gamma \Big(\frac{3}{2}, \frac{r^2}{4 \theta}\Big).\label{12}
	\end{equation}
	
Solution to Eq. (\ref{11}) cannot be obtained in closed form. However, numerically the value of $r_H$ can be found from the intersection of $e^{\nu}$ or $e^{-\lambda}$ with the $r/\sqrt{\theta}$ for minimal mass, as shown in Figs. \ref{fnu} and \ref{flam} respectively. For the range $r_H^2/4{\theta}>>1$, the lower incomplete gamma function can be expanded in the form provided in Eq. ({\ref{eq2}}) by iteration. Keeping up to first order term, we find 
\begin{equation}
	r_H \simeq \frac{2M}{\overline{Ric}}\Big(1-\frac{2M}{\sqrt{\pi \theta}}e^{- M^2/\theta}\Big).\label{13}
\end{equation}
	
There are different approaches also available in introducing  of noncommutativity in curved spacetime metric in Refs.~\cite{Lopez,Mukherjee,Chaichian}.
	
We can readily rewrite the mass as a function of $r_H$:
	\begin{equation}
	M=\frac{r_H \overline{Ric} \sqrt{\pi}}{4\gamma \Big(\frac{3}{2}, \frac{r^2}{4 \theta}\Big)}.\label{14}
	\end{equation}
	
	\begin{figure}[htp!]
		\includegraphics[scale=0.35]{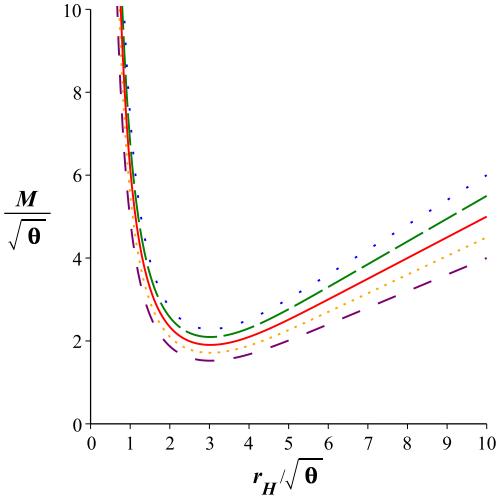}
		\caption{Variation of $M/\sqrt{\theta}$ vs $r_H / \sqrt{\theta}$ for different values of $\overline{Ric}$.}
		\label{fm}
	\end{figure}
	
The variation of the total mass is shown in Fig. \ref{fm}. The minimal mass $M_o$ for different values of $\overline{Ric}$ is explicitly shown in Fig. \ref{fm}, which increases with the $\overline{Ric}$. This minimal mass also confirms the degenerate horizon obtained in Figs. \ref{fnu} and \ref{flam}.
	
For the static noncommutative black holes, the Hawking temperature is defined as
\begin{eqnarray}
	T_H & &= \frac{1}{4 \pi}\frac{de^{\nu}}{dr}\sqrt{e^{\nu} e^{\lambda}}\Bigg|_{r=r_H}\nonumber\\
	& &=\frac{1}{4 \pi r_H \sqrt{Ric}} \Big(1-\frac{r^3_He^{-r_H^2/4\theta} }{4 \theta^{3/2}\gamma(\frac{3}{2}, \frac{r^2}{4 \theta})}\Big).\label{15}
	\end{eqnarray}
	
For large regime ($r^2/4\theta>>1$), one can easily recover the standard form of the Hawking temperature, given by
\begin{equation}
	T_H=\frac{1}{4 \pi r_H \sqrt{Ric}}.\label{16}
	\end{equation}  
	
\begin{figure}[htp!]
	\includegraphics[scale=0.35]{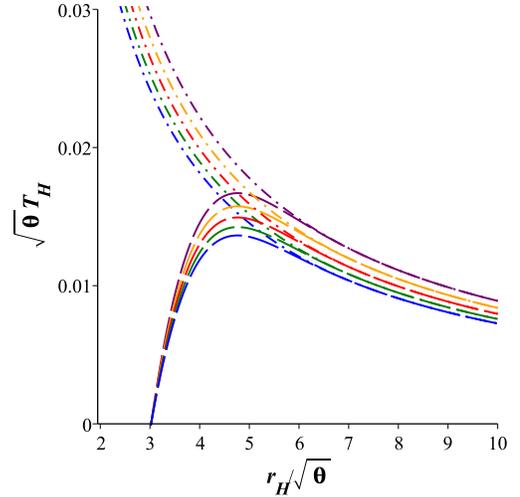}
	\caption{Variation of $T_H$ vs $r_H$ for different values of $\overline{Ric}$. Here, dashdot and longdash linestyle represent commutative and noncommutative spacetime, respectively.}
	\label{ft}
\end{figure}

\begin{figure}[htp!]
	\includegraphics[scale=0.35]{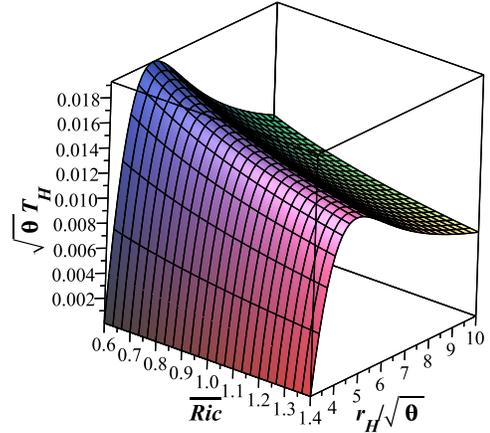}
	\caption{Variation of $T_H$ with $\overline{Ric}$ for the parametric values of $r_H/\sqrt{\theta}$.}
	\label{ftric}
\end{figure}

From Fig. \ref{ft}, it is clear that in the commutative consideration, the Hawking temperature diverges, which imposes a limit on the standard description of the Hawking radiation. For the region $r<r_o$, there is no black hole, therefore it is impossible to define temperature in that region. At the initial stage, $T_H$ reaches maximum from $r_H=r_o$ instead of exploding with $r_H$ up to the $r_H=4.8{\theta}$, and for $r_H>>r_o$, the temperature is the same as of the standard Hawking temperature. The Hawking temperature decreases with increasing $\overline{Ric}$. Moreover, the black hole remnant in higher $\overline{Ric}$ has a higher mass. The variation of temperature with $\overline{Ric}$ shown in Fig. \ref{ftric}.

The thermodynamic details of the quantum gravity system in the scenario of noncommutative geometry investigated respect to heat capacity. It can be obtained by the relation 
	\begin{equation}
	C=\frac{\partial M}{\partial r_H} \Big(\frac{\partial T_H}{\partial r_H}\Big)^{-1}.\label{17}
	\end{equation}
	
\begin{figure}[htp!]
		\includegraphics[scale=0.35]{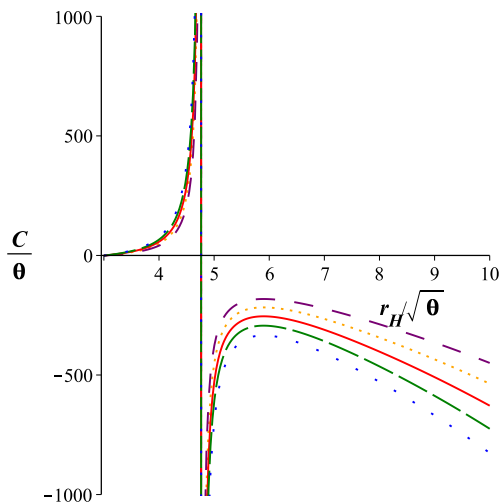}
		\caption{Plot of $ C/\theta$ vs $r_H/\sqrt{\theta}$ for different values of $\overline{Ric}$.}
		\label{fsp}
	\end{figure}

The behavior of the heat capacity is shown in Fig. \ref{fsp}. The heat capacity is positive for $r_o < r_H <r'$, which defines small and large black holes are stable. The black holes became unstable for $r' < r_H$ due to negative heat capacity. As $r_H$ goes to $r_o$, the heat capacity approaches to zero. For lower $\overline{Ric}$, the heat capacity is tending to positive, which also defines small black holes are more stable. 

From the first law of thermodynamics, the entropy of a black hole can be defined as $TdS=dM$. Hence	
	\begin{eqnarray}
		 \frac{dS_H}{dr_H} =\frac{\pi^{3/2} \overline{Ric}^{3/2}r_H}{\gamma(\frac{3}{2}, \frac{r^2}{4 \theta})}. \label{18}
	\end{eqnarray}
	
\begin{figure}[htp!]
		\includegraphics[scale=0.35]{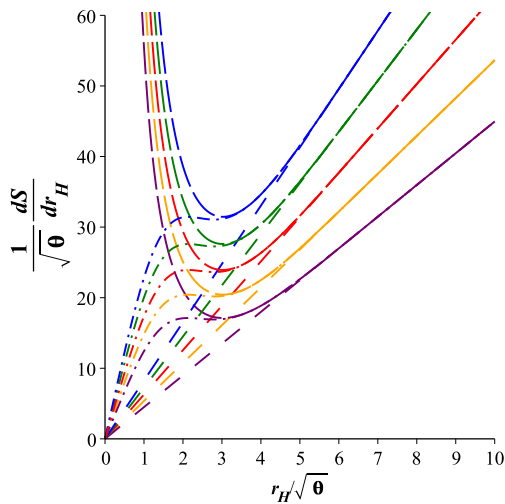}
		\caption{Variation of $\frac{1}{\sqrt{\theta}}\frac{dS}{dr}$ vs $r_H/\sqrt{\theta}$ for the different values of $\overline{Ric}$. Here longdash, dashdot and dash linestyle represent noncommutative spacetime, second-order correction of $\frac{dS}{dr}$ and semiclassical limit, respectively.}
		\label{fent}
	\end{figure}

The variation is shown in Fig. \ref{fent}. The Eq. (\ref{18}) can be expanded with the help of Eq. (\ref{eq2}).
	
The stability can also be examined by considering the free energy which can be defined as 
	\begin{equation}
	F= M - T_HS. \label{19}
	\end{equation}

	\begin{figure}[htp!]
		\includegraphics[scale=0.35]{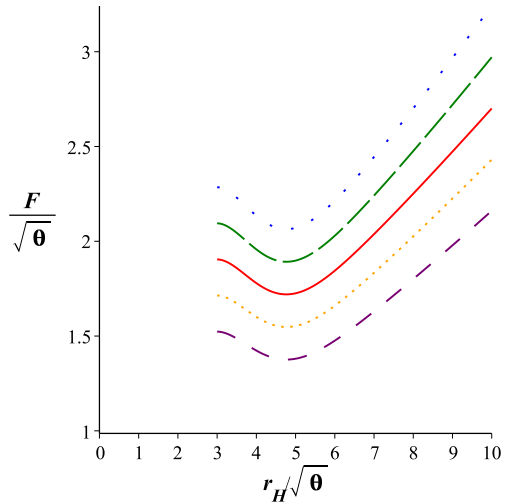}
		\caption{Variation of $F $ vs $r_H$ for different values of $\overline{Ric}$.}
		\label{feng}
	\end{figure}

The numerical calculation of the quantity is shown in Fig. \ref{feng}. Positive behavior of free energy supports the stability of a black hole. It is clear from the graph that the free energy becomes equivalent to minimum mass at $r=r_o$, thereafter decreases up to $r_+$ (corresponds to the maximum value of $T_H$) and then starts to increases.

\section{Conclusion}
In this paper, we have investigated the role of the noncommutative black hole in the framework of Finslerian spacetime after defining the radial coordinate and the total mass as $r'=r/\sqrt{\theta}$ and $M'=M/\sqrt{\theta}$. 
	
In our analysis, we have obtained the relation of $\overline{Ric}$ with the metric potential, temperature, entropy, heat capacity and free energy. By considering the particular Gaussian form of density, we observed that the minimal horizon ($r_o$) is always at $\simeq 3 \sqrt{\theta}$, for the different minimal masses ($M_o$) corresponding to different $\overline{Ric}$. Due to the negative pressure of the smeared object, it can look similar to the cosmological constant in the de Sitter universe. In fact, the line element near the origin supports the de Sitter-like behavior inside the inner horizon $r_c$.  It is found that $M_o$ varies with $\overline{Ric}$, in order of 0.191$ \sqrt{\theta}$, which is also confirmed from Fig. \ref{fm}. Therefore, the higher mass accretion can easily be explained in the Finslerian background. The Hawking temperature increases with decreasing of the $\overline{Ric}$, though the remnant masses are higher. As the horizon radius reaches  minimal horizon, the temperature of the noncommutative black hole vanishes. The temperature ($T_H$) varies with $\overline{Ric}$ in a nonlinear manner (vide Fig. \ref{ftric}). Therefore, we can expect that the micro-black holes with large fundamental energy-scale for higher $\overline{Ric}$ will be hotter, and at the endpoint of evaporation will have a smaller mass. From the variation of $\frac{dS}{dr_H}$ vs $ r_H/\sqrt{\theta} $, it is found that for second order correction the variation met at $\approx 3\sqrt{\theta}$, whereas the semiclassical approach met at $\approx 4.8 \sqrt{\theta}$, for the entire range of the Finsler parameter. The decrement of free energy from the minimum mass ($M_o$) means the evaporation decelerate in the vicinity of the zero temperature configuration.
	
One would consider that the weak energy conditions, $\rho_{\theta}+p_r \geq 0$ and  $\rho_{\theta}+p_t \geq 0$ are always satisfied. It is interesting to note that the strong energy condition  $\rho_{\theta}+p_r+2p_t \geq 0$ is not always satisfied. For the region $r<2\sqrt{\theta}$ strong energy does not obey. This implies that due to the dominance of quantum effects in the region under consideration, in spite of the nonlinear gravity, the classical description of energy and matter breaks down.
	
Due to the high temperature at the final stage of black hole evaporation, the effects of back reaction cannot be neglected in commutative background~\cite{Balbinot1,Balbinot2,Balbinot3}. In the noncommutative scenario, the temperature of the final stage is cool enough. Therefore, we can assume that the effects of back reaction should be suppressed. It is observed that the higher $\overline{Ric}$ reduces the temperature, as a consequence back reaction effects suppressed more intensively and less amount of energy released.
	
Finally, we can enunciate that the Finslerian parameter provides a range over which the noncommutative black hole can be analyzed and valid, and the higher order mass accretion of the system can easily be explained with the Finslerian background.

\appendix
	
\section{gamma function}
Gamma function can be defined as:
	\begin{equation}
	\Gamma(a) =\gamma(s,x)+\Gamma(s,x),
	\end{equation}
where $\gamma$(s,x) know as lower incomplete gamma function and $\Gamma$(s,x) as upper incomplete gamma function.
	
The lower incomplete gamma function is given by
	\begin{equation}
	\gamma (s, x)= \int_{0}^{x} t^{s-1}e^{-t} dt. \label{eq1}
	\end{equation}
	
	The upper incomplete gamma function is given by
	\begin{equation}
	\Gamma (s, x)= \int_{x}^{\infty} t^{s-1}e^{-t} dt.
	\end{equation}
	
For $x>>1$, the lower incomplete gamma function can expanded asymptotically as follows
	\begin{eqnarray}
	\gamma(s,x)& &= \Gamma(a)-\Gamma(s,x)\nonumber \\ 
	& &\simeq \frac{\sqrt{\pi}}{2}\Big[1-e^{-x} \sum_{\text{n=0}}^{+\infty}{\frac{x^{(1-2n)/2}}{\Gamma(\frac{3}{2}-n)}} \Big]. \label{eq2}
	\end{eqnarray}

\section*{ACKNOWLEDGMENTS}
SR and FR are thankful to the Inter-University Centre for Astronomy and Astrophysics (IUCAA), Pune, India for providing Visiting Associateship under which a part of this work was carried out. FR is also grateful to Jadavpur University for financial support under RUSA 2.0 and to DST-SERB (EMR/2016/000193), Government of India. SRC and DD are thankful to Debabrata Ghorai for several fruitful discussions related to the present work.


\begin{thebibliography}{99}
		
		\bibitem{hawking1} S. W. Hawking, Nature \textbf{248}, 30 (1974).
		
		\bibitem{hawking2} S. W. Hawking, Commun. Math. Phys. \textbf{43}, 199 (1975).
		
		\bibitem{Bekenstein1} J. D. Bekenstein, Phys. Rev. D \textbf{7} , 2333 (1973).
		
		\bibitem{Bekenstein2} J. D. Bekenstein, Phys. Rev. D \textbf{9}, 3292 (1974).
		
		\bibitem{Aharonov} Y. Aharonov, A. Casher, and S. Nussinov, Phys. Lett. B \textbf{191}, 51 (1987).
		
		\bibitem{Susskind1} L. Susskind and L. Thorlacius, Nucl. Phys. B \textbf{382}, 123 (1992).
		
		\bibitem{Susskind2} L. Susskind, Phys. Rev. Lett. \textbf{71}, 2367 (1993).
		
		\bibitem{Witten} E. Witten, Nucl. Phys. B \textbf{460}, 335 (1996).
		
		\bibitem{Seiberg} N. Seiberg and E. Witten, JHEP \textbf{1999}, 032 (1999).
		
		\bibitem{Snyder} H. S. Snyder, Phys. Rev. \textbf{71}, 38 (1947).
		
		\bibitem{Gruppuso} A. Gruppuso, J. Phys. A: Math. Gen. \textbf{38}, 2039 (2005).
		
		\bibitem{Nicolini1} P. Nicolini, J. Phys. A: Math. Gen. \textbf{38}, L631 (2005).
		
		\bibitem{Smailagic1} A. Smailagic and E. Spallucci, J. Phys. A: Math. Gen. \textbf{36}, L467 (2003).
		
		\bibitem{Smailagic2} A. Smailagic and E. Spallucci, J. Phys. A: Math. Gen. \textbf{36}, L517 (2003).
		
		\bibitem{Chaichian} M. Chaichian, A. Demichev, and P. Presnajder, Nucl. Phys. B \textbf{567}, 360 (2000).
		
		\bibitem{Cho} S. Cho, R. Hinterding, J. Madore, and H. Steinacker, Int. J. Mod. Phys. D \textbf{9}, 161 (2000).
		
		\bibitem{Smailagic3} A. Smailagic and E. Spallucci, J. Phys. A: Math. Gen. \textbf{37}, 7169(2004).
		
		\bibitem{Nicolini2} P. Nicolini, A. Smailagic, and E. Spallucci, Phys. Lett. B \textbf{632}, 547 (2006).
		
		\bibitem{Myung} Y. S. Myung, Y. W. Kim, and Y. J. Park, Phys. Lett. B \textbf{656} 221 (2007). 
		
		\bibitem{Banerjee1} R. Banerjee, B. R. Majhi, and S. Samanta, Phys. Rev. D \textbf{77}, 124035 (2008).
		
		\bibitem{Banerjee2} R. Banerjee, B. R. Majhi, and S. K. Modak, Class. Quant. Gravit. \textbf{26}, 085010 (2009).
		
		\bibitem{Garattini} R. Garattini and F.S.N. Lobo, Phys. Lett. B \textbf{671}, 146 (2009).
		
		\bibitem{Lobo} F. S. N. Lobo and R. Garattini, JHEP \textbf{1312}, 065 (2013).
		
		\bibitem{Kuhfittig} P. K. F. Kuhfittig, Adv. High Energy Phys. \textbf{2012}, 462493	(2012).
		
		\bibitem{Rahaman3} F. Rahaman, S. Islam, P. K. F. Kuhfittig, and S. Ray, Phys. Rev. D \textbf{86}, 106010 (2012).
		
		\bibitem{Rahaman1} F. Rahaman, P. K. F. Kuhfittig, B. C. Bhui, M. Rahaman, S. Ray, and U. F. Mondal, Phys. Rev. D \textbf{87}, 084014 (2013).
		
		\bibitem{Rahaman2} F. Rahaman, P. K. F. Kuhfittig, K. Chakraborty, A. A. Usmani, and S. Ray, Gen. Relativ. Gravit.  \textbf{44}, 905	(2012).
		
		\bibitem{Radinschi} I. Radinschi, F. Rahaman, and U. F. Mondal, Int. J. Theor.	Phys. \textbf{52}, 96 (2013).
		
		\bibitem{Kim} W. Kim, E. J. Son, and M. Yoon, JHEP \textbf{2008}, 042 (2008).
		
		\bibitem{Bao2000} D. Bao, S. S. Shern, and Z. Chen, \emph{An introduction to Riemann-Finsler geometry, Graduate Texts in Mathematics} (Springer, New York, 2000).
		
		\bibitem{Pfeifer2014} C. Pfeifer and M. Wohlfarth, In: Bi{\v c}{\'a}k J., Ledvinka T. (eds) Relativity and Gravitation, Springer Proceedings in Physics {\bf157} (Springer, Cham, 2014).
		
		\bibitem{Kouretsis2012} A. P. Kouretsis, M. Stathakopoulos, and P. C. Stavrinos, Proceedings of M3ST2012 (2012).
		
		\bibitem{Finsler1918} P. Finsler, {\"U}ber Kurven und Fl{\"a}chen in allgemeinen R{\"a}umen, Ph.D. Thesis, Georg-August Universit{\"a}t zu G{\"o}ttingen (1918).
		
		\bibitem{Caponio1} E. Caponio, M. A. Javaloyes, and A. Masiello, Math. Ann, \textbf{351}, 365 (2011).
		
		\bibitem{Caponio2} E. Caponio, M. A. Javaloyes, and M. Sanchez, Rev. Mat. Iberoamericana, \textbf{27}, 919 (2011).
		
		\bibitem{Caponio3} E. Caponio, M. A. Javaloyes, and A. Masiello, J. Phys. A: Math. Theor. \textbf{43}, 135207-135222 (2010).
		
		\bibitem{Hohmann2017} M. Hohmann and C. Pfeifer, Phys. Rev. D {\bf95}, 104021 (2017).
		
		\bibitem{Girelli2007} F. Girelli, S. Liberati, and L. Sindoni, Phys. Rev. D {\bf75}, 064015 (2007).
		
		\bibitem{Gibbons2007} G. W. Gibbons, J. Gomis, and C. N. Pope, Phys. Rev. D {\bf76}, 081701 (2007).
		
		\bibitem{Minguzzi2015} E. Minguzzi, Class. Quantum Grav. {\bf32}, 185008 (2015).
		
		\bibitem{Stavrinos2018} P. C. Stavrinos and M. Alexiou,  Int. J. Geo. Meth. Mod. Phys. {\bf15}, 1850039 (2018).
		
		\bibitem{Aazami} A. B. Aazami and M. A. Javaloyes, Class. Quantum Gravit. \textbf{33}, 025003 (2016).
		
		\bibitem{Caponio4} E. Caponio and G. Stancarone, Class. Quantum Gravit. \textbf{35}, 085007 (2018).
		
		\bibitem{Caponio5} E. Caponio and G. Stancarone, Int. J. Geo. Meth. Mod. Phys. \textbf{ 13}, 1650040 (2016).
		
		\bibitem{Li2014} X. Li and Z. Chang, Phys. Rev. D {\bf90}, 064049 (2014).
		
		\bibitem{Lopez} J. C. Lopez-Dominguez, O. Obregon, M. Sabido, and C. Ramirez, Phys. Rev. D \textbf{74}, 084024 (2006).
		
		\bibitem{Mukherjee} P. Mukherjee and A. Saha, Phys. Rev. D \textbf{77}, 064014 (2008).
		
		\bibitem{Chaichian} M. Chaichian, A. Tureanu, and G. Zet, Phys. Lett. B \textbf{660}, 573 (2008). 
		
		\bibitem{Balbinot1} R. Balbinot and A. Barletta, Class. Quantum Gravit. \textbf{6}, 195 (1989).
		
		\bibitem{Balbinot2} R. Balbinot, A. Fabbri, V. Frolov, P. Nicolini, P. J. Sutton, and A. Zelnikov, Phys. Rev. D \textbf{63}, 084029 (2001).
		
		\bibitem{Balbinot3} R. Balbinot, A. Fabbri, P. Nicolini, and P. J. Sutton, Phys. Rev. D \textbf{66}, 024014 (2002).
		
	\end{thebibliography}
\end{document}